# Calculating Vibronic Spectra with a linear algorithm based on Gaussian Boson Sampling

Iurii Konyshev[1], Ravi Pradip[1], Oliver Page[1], Caghan Ünlüer[1], Rinat T. Nasibullin[2], Vladimir V. Rybkin[3], Wolfram Pernice[1,*], Simone Ferrari[1]

[1] University of Heidelberg, Kirchhoff-Institute for Physics, Im Neuenheimer Feld 227, 69120 Heidelberg, Germany
[2] University of Helsinki, Department of Chemistry, Faculty of Science, P.O. Box 55 (A.I. Virtanens plats 1), FIN-00014, Finland
[3] HQS Quantum Simulations GmbH, Rintheimer Str. 23, 76131 Karlsruhe, Germany

* Correspondence to: wolfram.pernice@kip.uni-heidelberg.de

**Accurately simulating molecular vibronic spectra remains computationally challenging due to the exponential scaling of required calculations. Here, we show that employing the linear coupling model within the gaussian boson sampling framework effectively addresses this limitation. We implement the algorithm for simulating the pentacene molecule through three distinct approaches, using numerical simulation on a classical computer and experimentally using two optical setups equipped with different photon detectors (SNSPD and SPAD). High fidelity (F > 0.999) was achieved between the simulated Franck–Condon profiles and analytically calculated profiles obtained by enumerating all possible transitions within the linear coupling model. Furthermore, simulations were performed for larger molecular systems using 48 vibrational modes of naphthalene and 64 vibrational modes of anthracene. Comparison with experimental data confirms that the simulated spectra accurately reproduce both the positions and shapes of the measured spectral bands. A notable advantage of our algorithm is its scalability, requiring only a fixed minimal set of optical components irrespective of the size of the studied system.**

# Introduction

The ability to rapidly compute the spectra of large molecules is crucial across various fields, including the development of new organic light-emitting diodes (OLEDs)[1], and solar cells[2]. In medicine and biological research, detailed knowledge of vibronic spectra enables the development of fluorescent probes and biomarkers, which are vital for advanced imaging techniques and diagnostics[3]. The vibronic transition probabilities between two electronic states are directly related to the Franck-Condon factors (FCFs)[4,5], which quantify the overlap of vibrational wavefunctions between different electronic states. Due to their critical importance in understanding chemical phenomena such as spectroscopy and photochemistry, numerous computational strategies have been developed and refined for calculating FCFs using classical computers[6–10].

The conceptually simplest algorithm is the time-independent sum-over-states (SOS) method. In practice, SOS enumerates every combination of vibrational quantum numbers up to a chosen cutoff, evaluates the corresponding Franck–Condon factor, and sums the resulting intensities[11]. Its Achilles' heel is combinatorial growth: once the mode count exceeds a few dozen, exact SOS with naive enumeration becomes computationally impractical[12].

This exponential scaling limits its practicality for larger molecular systems, driving the search for more efficient approaches.[11,13] In 2015, Huh and colleagues proposed an algorithm for a quantum simulator that addresses this problem by enabling the calculation of vibronic spectra for large molecules using quantum computing techniques[14]. The authors introduced a method in which the vibronic band profile is considered within the Franck-Condon approximation while accounting for the Duschinsky effect—a phenomenon involving the mixing of vibrational modes between electronic states[15]. By mapping the calculation of vibronic spectra onto a boson sampling problem, they demonstrated that a quantum simulator could perform these computations more efficiently than classical methods. This approach was called vibronic boson sampling (VBS). In VBS, the number of normal vibrations of the molecule is equal to the number of optical modes of the simulator. Thus, the hardware requirements for such a computing machine grow linearly with the size of the system under study.

The realization of this algorithm (in whole or in part) was demonstrated in various platforms such as superconducting qubits[16], trapped ions[17,18], bulk[19] and integrated optics[20]. Despite these advancements, most experimental realizations have so far been limited to the vibronic spectra of two-mode molecular systems. It is also worth highlighting the work[21], which successfully demonstrated the calculation of a four-mode system. This achievement currently holds the record for the largest system calculated using the VBS approach. As the authors point out, the primary challenge in implementing the algorithm on integrated photonic microprocessor chips lies in the constraints imposed by the squeezing level and the broadband spectral characteristics of the squeezed light source. These limitations significantly reduce the interference visibility between the coherent light and the squeezed light[20,21]. A key challenge is the simultaneous realization of the displacement operator and the squeeze and rotation operators on the chip. Further scaling leads to significant technical challenges, such as maintaining precise phase settings due to heating effects and thermal crosstalk. Achieving a substantial leap to systems with qualitatively different sizes (more than 30–40 modes) thus requires the development of fundamentally new methodologies.

Alternatively, by neglecting the Duschinsky effect and squeezing, i.e. assuming the same vibrational frequencies in the initial and final electronic states, the linear coupling model (LCM) can be adopted as the basis of a quantum simulator. Computational methods built on the LCM model are widely used to calculate vibronic spectra and can reliably reproduce experimental results for complex molecular systems[13,22–25]. In particular, integrated electronic structure calculations with vibronic modeling have been used to predict the optical properties of novel chromophores[26]. By employing robust approximations analogous to LCM, this approach not only confirmed the accuracy of such methods for describing vibronic spectra but also guided the rational design of new mechanochromic and luminescent materials.

Here, we present a method leveraging an optical setup consisting of a tunable laser source, a controlled optical attenuator, and a single photon detector to sequentially simulate and analyze molecular vibrational modes. The fact that in the LCM model harmonic oscillators can be considered separately from each other gives us the ability to compute a system of any size using only one optical mode. By initially calibrating the experimental apparatus, this approach efficiently mitigates the adverse effects of optical losses, thereby ensuring high fidelity in the resulting vibrational spectra simulations. Our method possesses intrinsic scalability, as the experimental complexity remains constant regardless of the molecule size, with only the experimental runtime increasing linearly with system size. This scalability makes the presented method highly suitable for computationally intensive applications. Additionally, we demonstrate the versatility of the algorithm through successful implementation on a classical computer, highlighting its broad accessibility for efficiently computing Franck–Condon profiles of large and complex molecular systems. The calculation throughput could be further enhanced by deploying multiple processing elements in parallel on a photonic chip, enabling simultaneous acquisition of statistics for multiple vibrational modes or multiple molecules.

# Results

## Mapping the LCM to photonics

Molecules absorb or emit light at frequencies determined by allowed transitions between different electronic states. These electronic transitions are accompanied by changes in the vibrational energy of the molecules. The term vibronic refers to the simultaneous vibrational and electronic transitions of a molecule upon absorption or emission of light. We focus on transitions between the zero vibrational level of one harmonic oscillator and the few lowest vibrational levels of another harmonic oscillator (Zero-Temperature Approximation). Under the harmonic and Franck–Condon (FC) approximations[27], the intensity of a vibronic transition is proportional to the square of the overlap integral between the wave functions of two multidimensional quantum harmonic oscillators (QHO). The square of the overlap integral is called the Franck–Condon factor (FC factor). The dimension of each QHO wave function is equal to the number of normal modes $N$. In the general case $N = 3M - 6(5)$, for nonlinear(linear) molecules with $M$ atoms. One multidimensional harmonic potential approximates the potential energy surface of the initial electronic state, and the other one describes the potential energy surface of the final electronic state of transition

In the linear coupling model (LCM) used here, we assume that for the two electronic states involved in the transition, the form and frequency of the normal modes remain the same. The only difference lies in the equilibrium positions around which the normal vibrations occur. In this scenario, because mixing of modes is absent, the molecular vibrations can be represented as a set of independent QHOs. Hence, the Franck-Condon factor is reduced to the product of the squares of the one-dimensional HO wave functions overlap integrals:

$$|\langle 0_1, 0_2, \ldots, 0_N \mid j_1, j_2, \ldots, j_N \rangle|^2 = \prod_{i=1}^{N} |\langle 0 \mid j_i \rangle|^2, \qquad (1)$$

where $j_i$ is vibrational quanta in the $i$th normal mode of one electronic state and another electronic state do not have vibrational quanta in for all $N$ modes. The one-dimensional overlap integrals in this case can be found using the expression:

$$|\langle 0 | j_i \rangle|^2 = \frac{S_i^j}{j!} e^{-S_i}, \qquad (2)$$

where $S_i$ is the Huang-Rhys (HR) factor of the $i$th mode[28]. The HR factors can be interpreted as the average number of vibrational quanta excited during the electronic transition and depends quadratically on the difference between the equilibrium positions of the harmonic oscillators ($\Delta Q_0$) (see Figure 1). Those normal modes for which there is no change in the position of the harmonic oscillator minimum during the electronic transition have a $\langle 0 | 0 \rangle = 1$ and therefore do not contribute to the vibronic progression. Conversely, the greater $\Delta Q_0$, the more likely it is that transitions occur with phonon excitation. As a result, the $0-0$ line becomes less intense, whereas the other lines in the spectrum gain on intensity. As a consequence, modes with a higher value of $S_i$ contribute more to the vibronic transition profile. The energy of a vibronic transition can be found using the expression:

$$E_{tr} = E_{0-0} \pm \sum_{i=1}^{N} E_i j_i, \qquad (3)$$

where $E_{0-0}$ is energy of electronic transition (zero phonon line), $E_i$ is energy of i-th QHO. The sign in this equation depends on the type of transition For absorption, the initial state is the ground state as shown in in Figure 1 and the 0-0 line has, therefore, the minimum energy value in the FC profile.

Knowing $E_i$ and $S_i$ for all $N$ harmonic oscillators and using eqs. (1) to (3), we can calculate the energy and FC factor for any vibrational configuration $|j_1, j_2, ..., j_N\rangle$. By systematically enumerating every such configuration—a procedure known as the time-independent sum-over-states (SOS)

approach—we recover the complete FC profile of the system. The difficulty here is that the number of available states is:

$$P = (1 + K)^N, \qquad (4)$$

where $K$ is maximal value of vibrational quanta. As the number $N$ increases, the number of calculations that must be performed to enumerate all possible states grows exponentially. We will refer to the spectra obtained by this method as reference spectra.

An alternative approach to calculating the profile is vibronic boson sampling. During the experiment we register events that we attribute to a particular state of the system, accumulate statistics of these events and then build a histogram to obtain an FC profile[14]. The advantage of this scheme is that it inherently takes into account states with a largest values of FC factors. An important parameter for this scheme is the total number of registered events $P$, this number specifies the accuracy $\epsilon$ with which we build the FC profile. For a fixed value of $P$, an increase in the number of modes of the molecule under study $N$ does not lead to a decrease in the calculation accuracy $\epsilon$[14,21].

It is easy to see from equation (2) that in LCM transition probabilities for one dimensional harmonic oscillator obey Poisson statistics, and the HR factor $S$ can be interpreted as the average number of phonons induced by an electronic transition. We also know that coherent states of light exhibit the same Poisson statistics in the Fock basis:

$$|\langle n \mid \alpha \rangle|^2 = \frac{\langle n \rangle^j}{j!} e^{-\langle n \rangle}, \qquad (5)$$

This is not a coincidence because a coherent state can be created by applying a displacement operator $\hat{D}(\alpha)$ to the vacuum state $|0\rangle$. Thus, measuring a coherent state in the Fock basis can be viewed analogously as a transition from the zero vibrational level of a displaced harmonic oscillator to the levels of a original harmonic oscillator. We observe that the HR factor $S$ and the average photon number $\langle n \rangle$ in a coherent state have similar physical meanings: $s$ represents the number of phonons created by an electronic transition in the molecular system, while $\langle n \rangle$ represents the number of photons created by displacement of the vacuum state in the optical system.

In order to encode a molecule's HR factors into the mean photon numbers $\langle n \rangle$ of optical pulses we use coherent laser light. By employing photon-number-resolving (PNR) detectors to measure in the Fock basis, we can then collect photon-number statistics. These observed statistics correspond directly to the to the statistics of transitions between energy levels of quantum harmonic oscillators that describe the molecule's normal vibrational modes. Knowing these statistics and values of energies of each harmonic oscillator allows us to construct a vibronic transition profile. Using a single photon detector and attenuated laser pulses, it is possible to model the vibronic spectrum of molecules without restrictions on the size of the system under study with an accuracy sufficient to interpret experimental data or make predictions. In Fig.2 we outline a step-by-step algorithm for boson sampling in the LCM which is used for the experimental study presented in the following:

1. Calculate the Huang-Rhys factor - $S_i$ and the energy - $E_i$ of each normal mode of the molecule under study (see Methods for details).

2. Initialize the optical setup using tunable attenuators such that the photon number statistics from the single photon detectors correspond to the Poisson statistics of coherent light with $\langle n \rangle_1 = S_1$. (Alternatively, a random number generator that follows Poisson statistics can be used. The main value for the generated distribution should be set to the Huang-Rhys factor of the chosen mode.)

3. Collect photon number statistics from the detector that correspond to the first normal mode of the molecule.

4. Repeat steps 2 and 3 $N-1$ times.

5. After step 4 we have $N$ arrays with photon number statistics that correspond to the $N$ vibrational normal modes of the molecule. Multiply each array by its corresponding energy $E_i$

6. Perform the summation over the $N$ arrays element by element.

7. Generate a histogram for the resulting array. This histogram is the FC profile for the molecule under study.

## Experimental validation using SPAD and SNSPD

To validate the algorithm we used the setup shown in Fig. 3, employing a coherent light source and a solid-state single-photon detector. The optical signal is generated using a calibrated and attenuated pulsed laser (*Pritel FFL-1550*), operating at a repetition rate of 5 MHz. The laser output is attenuated with a variable optical attenuator (*VOA, HP8156A*) providing a range of 0 to 60 dB. Afterwards the signal is split into two equal parts using a 50:50 optical splitter (S). One output of the splitter is directed to a calibrated photodiode (*PD, Thorlabs S154C*), which continuously monitors the average power at the attenuator output. The other output passes through a fixed attenuator (A) providing an additional 60 dB before being sent to a solid-state single-photon detector (*InGaAs SPAD, Princeton Lightwave PGA600*). To reduce dark counts and afterpulsing, the detector is operated in gated mode. This configuration ensures precise control of the average photon number per pulse while maintaining continuous monitoring of laser power fluctuations. The laser includes an internal fast photodiode that generates a synchronization signal, which is conditioned to match the detector's input requirements and to account for optical path delays. A pulse generator (*HP8131A*), triggered by the laser's synchronization signal, performs this conditioning and provides a start pulse for the time tagger unit.

The time tagger unit (*Swabian Instruments Time Tagger Ultra*) records timestamp streams from both the laser trigger and the detector. Absence of a detection event, corresponding to zero photons, is represented as a "0" in the output array, while a detection event produces a timestamp recorded as a "1." The resulting binary array reflects the Poisson distribution of the input photon number, where the "1"s include contributions from events with higher photon numbers. The average photon number per pulse is calculated by analyzing the fraction of zeros in the array, factoring in the detector efficiency at the applied bias voltage. The system's accuracy is verified by confirming that the output data follows a Poisson distribution.

To determine the vibronic spectrum, attenuation levels on the VOA are set to correspond to the Huang–Rhys factor $S_i$ for specific normal modes of the molecule. The timestamps from the detector are collected to construct an output array for $i$th normal mode of the molecule. We then repeat this procedure for the Huang- Rhys factors corresponding to other vibrational modes. The spectrum is obtained by multiplying each array by its corresponding energy $E_i$ and summing the arrays element-wise. A histogram of the resulting array is then generated to visualize the vibronic spectrum.

To further validate the proposed algorithm with alternative single-photon detection technologies, the same experiment was repeated using a superconducting nanowire single-photon detector (*SNSPD*). The optical configuration—comprising a calibrated and attenuated pulsed laser, operating at 40 MHz, variable optical attenuator (*HP8156A*), optical splitter (50:50), and photodiode (*Thorlabs S154C*) for continuous power monitoring—remained unchanged. The SNSPD was selected for its superior performance compared to the InGaAs SPAD, offering negligible afterpulsing and significantly lower dark count rates ($< 75Hz$). These advantages also eliminated the need for the signal-conditioning pulse generator used in the SPAD setup. The SNSPD, integrated on a silicon nitride-on-insulator waveguide platform[29,30], was cooled to 1.6 K in a closed-cycle helium cryostat (*Janis Research*). The optical output was coupled to the SNSPD via an on-chip apodized grating coupler. On the electrical side, RF cabling carried the photon detection signal from the cryostat to an external bias T (*Mini-Circuits ZFBT-6GW+*). The DC port of the bias T was connected to a Keithley 2450 SourceMeter for biasing the SNSPD, while the RF signal was amplified using two cascaded low-noise RF amplifiers (*Mini-Circuits ZFL-1000LN+*). The amplified response of the SNSPD was then fed into a time tagger unit (*Swabian Instruments Time Tagger X*) for precise timestamp recording and processing under identical conditions to the SPAD measurement.

We use the experimental setup in Figure 3 to generate Poisson-distributed numbers matching the Huang–Rhys factors ($S$). However, the same arrays can be generated entirely in software using Python's NumPy library. This purely computational approach has the benefit of requiring no specialized equipment, but it raises the question of whether pseudo-random number generation is sufficient for accurately computing vibronic spectra. Since each new pseudo-random number depends on previous outputs, we need to ensure that any resulting correlations do not significantly affect the calculated Franck–Condon profile.

## Experimental results for the pentacene molecule

We chose the pentacene molecule as the test object for our algorithm. Pentacene is a rigid, flat molecule that consists of five conjugated carbon rings (see Fig.4(a)), which is suitable for the LCM[13,26]. The electron dipole transition $S_0 - S_1$ is strong, which allows using the FC approximation[25]. Since the geometry of this molecule changes slightly during the transition $S_0 - S_1$, its normal modes of have small values of HR factors $S < 0.25$. This allows us to limit our study to transitions between levels 0 and 1 of harmonic oscillators because the probabilities of transitions between other states are relatively small[31]. In this case we can use threshold single photon detector instead of PNR detectors. Since the molecule is well studied, one can find both calculation results[11,25] and results of high-resolution spectra obtained experimentally at 5 K[32].

In order to understand how the imperfection of hardware affects the operation of the algorithm, we conducted an analytical calculation for the reference FC profile. Using (eqs. (1) to (3)), we determined the FC factor for all possible states of our system, limiting the number of modes to eight and the maximum vibrational quanta for each oscillator to 1. This means that the total number of possible states in this case is equal $2^8 = 256$. We used the energy values $E$ for harmonic oscillators and their HR factors $S$ from[31]. To quantitatively compare the reference profile and the profiles obtained by our algorithm, we defined Fidelity $F = \sum_i p_i q_i$ where $p$ and $q$ are the normalized reference and simulated spectra[21].

Under ideal conditions, as the number of events registered by the simulator increases, we detect an increasing number of unlikely events and the profile from the simulation will converge to the reference spectrum with $F \to 1$. As shown in Fig.4(b), we are limited by the imperfection of our setup, thus after a certain number of events the fidelity saturates. The maximum fidelity for the measurements using the SPAD is 0.997, which is reached within 100 000 counts. The maximum fidelity for SNSPD is 0.9997 which is reached within 1 000 000 counts, demonstrating a more accurate result than with SPAD. When using the NumPy script we do not find any limits for maximum fidelity. For the maximum value of counts $P = 100000000$ that we are investigated we obtain $F = 0.9999996$. Even though the NumPy script yields more accurate results than the optical implementation options, the difference between these results is small and already at $P = 10000$ we have $F = 0.99$ in all cases. Hence, the algorithm generates a spectrum very close to the reference when using an optical setup with SNSPD or the NumPy library. The key advantage of this approach is that we no longer have to exhaustively enumerate the entire vibrational state space, whose size grows exponentially with the number of modes.

In Figure 4(c), we compare high-resolution absorption spectrum of pentacene measured in a hexadecane matrix at 5 K with the reference Franck–Condon profile calculated analytically and the results of simulations using both detectors with 10 000 events. The theoretical intensities were broadened using Lorentzian functions with a full width at half maximum (FWHM) of approximately $30 cm^{-1}$. Overall, the theoretical spectrum shows good agreement with the experimental data, successfully reproducing even low-intensity spectral features. The higher-energy vibronic transitions predicted by the model (beginning at around 1000 $cm^{-1}$) appear to be slightly shifted (around 100 $cm^{-1}$) toward higher frequencies compared to the experimental values, potentially due to interactions between the Pentacene molecules and the host matrix. Grimme et al.[25] observed a similar shift when comparing their calculation results with these experimental data.

Despite the fact that the results obtained using SPAD differ more pronounced from the reference spectrum, which indicates a greater influence of errors introduced by the experiment compared to other simulations, these results agree best with the experimental data (see Fig. 4(c and d)). This indicates that for this particular case, further increasing the number of events is inappropriate since it will lead to the spectra obtained from simulations tending to the reference. The influence of errors introduced by the optical equipment is less than the influence of the approximations we use for calculation. Thus, we conclude that the proposed algorithm for simulating the FC profile based on the LCM and FC approximation predicts the absorption spectrum of the vibronic band of the Pentacene molecule with good accuracy for all three platforms already at 10,000 events. The ability to implement the algorithm with high accuracy using photonics opens up further opportunities for improving its processing speed, since we can add electronics to our optical circuits that will process the data coming from the detectors and ensure the execution of steps 5 and 6 of the algorithms at the hardware level. Integrated photonics would further enable scaling of this experiment by incorporating dozens of optical modes, each equipped with a tunable attenuator and detector, onto a single chip. This facilitates parallel computation for multiple molecules at once.

**Simulation of naphthalene, anthracene and pentacene molecules.**

To confirm that the algorithm also works in the case when the maximum value of the vibrational quantum $K$ is higher than 1, we performed the same calculation of fidelity that we made for 8 modes of pentacene for naphthalene and anthracene. These molecules possess two and three conjugated carbon rings, respectively (see Fig.6(a,b)). We chose these molecules because they have properties similar to pentacene, yet the maximum HR value is higher (about 0.33 and 0.45, respectively), while for pentacene this value is 0.25. For these molecules, we used a maximal value of vibrational quanta $K = 3$ and took into account all modes that have HR factors $S > 10^{-5}$ (9 for naphthalene and 12 for anthracene). To demonstrate that significantly increasing the number of vibrational modes does not require accumulating a proportionally larger number of events to maintain comparable fidelity, calculations were performed for pentacene, expanding the number of vibrational modes considered from 8 to 18. As in the previous case $K = 1$. This example is illustrative because increasing the number of vibrational modes drastically raises the number of states computed for the reference spectrum—from 254 to 262,000—which inevitably results in a higher number of events needed to approach unity fidelity. The results of these calculations are shown in Fig.5. All three curves show similar behavior. The fidelity values approach unity at slightly different rates; however, for the same number of events, the average fidelity values and their standard deviations remain comparable. For event counts exceeding 100,000, the fidelity consistently surpasses 0.99 across all studied molecules. This result confirms that the algorithm functions robustly for different values of $K$ and is not restricted to the case of K=1. Specifically, comparing the results for pentacene, we observe that increasing the number of vibrational modes from 8 (see Fig. 4) to 18, while maintaining the same event count, decreases the fidelity slightly from 0.997 to 0.991. Nonetheless, as previously demonstrated with pentacene (Fig. 4), a fidelity of approximately 0.99 is more than sufficient for accurately comparing or predicting experimental data.

## Discussion

Although the algorithm presented in this paper employs a zero-temperature approximation, we compare our results with absorption spectra measured at room temperature. This choice is justified by practical considerations, as information regarding the optical properties of molecules under ambient conditions is typically of greater relevance for most applications. For naphthalene and anthracene, a comparison of the experimental data[32,33] with our simulations results for 100,000 events is shown in the Fig.6. In this simulation, we took into account all 66 modes for anthracene and 48 modes for naphthalene. We also did not set the limit for the maximum value of vibrational quanta $K$. All spectra are normalized to the maximum intensity value and expanded using Gaussian functions with a full width at half maximum, $100 cm^{-1}$ (dashed line), $300 cm^{-1}$,$190 cm^{-1}$ (solid red line on a) and b) respectively).

In both cases, the simulations (solid green lines in Fig.6) closely reproduce the experimental peak positions and shapes. The only appreciable difference concerns the relative intensities of the side-bands: the computed amplitudes attenuate more rapidly than the experimental ones, a minor effect for naphthalene but a conspicuous one for anthracene. An analogous intensity shortfall was reported by Benkyi *et al.*[34], who examined anthracene with three successively refined models: the linear-coupling model (LCM), an extension that includes mode-frequency shifts (the parallel approximation), and a variant that further incorporates the full

Duschinsky rotation. While each approach faithfully reproduced the overall spectral envelope, all of them systematically underestimated the vibronic-band intensities relative to the 0–0 origin. Hence, adjusting the vibrational frequencies and accounting for Duschinsky mixing in this case does not bring theory into closer agreement with the measurements. The missing intensity was instead ascribed to Herzberg–Teller (HT) vibronic coupling, a mechanism that allows the transition dipole moment to vary linearly with nuclear displacements[35]. Independent calculations by Santoro and co-workers confirmed that explicitly including HT terms restores the strength of the higher-energy side-bands and aligns the computed spectrum with experiment[36].

Our computational scheme, like the original concept of vibronic boson sampling[14], is formulated entirely within the Franck–Condon (FC) approximation, which treats the electronic transition-dipole moment as independent of nuclear displacements[27]. Because Herzberg–Teller (HT) coupling lies outside this approximation, all intensity-borrowing pathways associated with a coordinate-dependent dipole are excluded by construction. Consequently, any residual mismatch between the simulated and experimental spectra arises from the FC assumption itself, rather than from the LCM or the algorithm employed.

We further used a Gaussian line broadening of $100\ cm^{-1}$ to demonstrate that the simulation can identify which lines contribute the most to the formation of the absorption band (see dashed green lines in Fig.6). Since the algorithm directly yields FC factors without intrinsic spectral broadening, it enables not only the accurate prediction of experimental spectra but also facilitates the detailed analysis of previously measured spectra. This result demonstrates that for molecules for which the LCM and FC approximation work well, the algorithm allows simulating FC profiles with an accuracy sufficient for predicting and analyzing experimental spectra. Without limiting the values of the maximum vibrational quantum and the number of modes of the system under study. This allows us to perform calculations for very large systems using standard compute hardware without spending significant time resources. (In our case, the calculation on a regular office laptop took less than 5 seconds.) An additional advantage of this approach is that it does not rely on a priori assumptions about which vibrational modes will contribute most significantly. Instead, the most probable transitions—which dominate the spectral intensity—are naturally accounted for within the simulation.

A particularly significant advantage of our method lies in the strongly reduced input-data demand. In the conventional vibronic boson-sampling scheme, one must compute the Hessians of both the ground and the excited electronic states. These tensors are needed to construct the Duschinsky rotation matrix and to extract the squeezing and displacement parameters that are subsequently encoded in the photonic device[14]. Obtaining the excited-state Hessian is exceptionally demanding and often becomes the principal bottleneck in vibronic-spectrum simulations[11]. The Linear Coupling Model (LCM) removes this limitation: it requires only the ground-state Hessian plus the excited-state energy gradients evaluated along the ground-state normal coordinates. This gradient is much easier to compute than the hessian of the excited state[11]. From these readily accessible quantities one can compute the Huang–Rhys factors and vibrational frequencies (see Methods for details). Here we show that this minimal data set is sufficient to simulate Franck–Condon band profiles that closely match experimental spectra for rigid molecules such as the polyacenes.

## Conclusion

We proposed an efficient algorithm for the calculation of Franck - Condon profiles, featuring linear growth in the number of operations with the growth of the number of modes of the system under study and the absence of restrictions on the value of the maximum vibrational quantum. We validated the performance experimentally with a photonic setup and computationally using an off-the-shelf computer using the example of the pentacene molecule. Comparison of the results of these simulations with the results of the analytical calculation showed that it is possible to achieve a high-fidelity value (>0.999) for both computing platforms. We also carried out a similar calculation on a laptop for the molecules of naphthalene and anthracene and obtained similar results.

For anthracene and naphthalene, we performed additional simulations in which all 66 and 48 vibrational modes, respectively, were included with no limit on the maximum vibrational quantum number. Comparison with room-temperature experimental spectra shows that the algorithm accurately reproduces the positions, shapes, and intensities of the vibronic bands of naphthalene. For anthracene, however, it systematically underestimates the vibronic-band intensities relative to the 0–0 transition. We attribute this to its reliance on the Franck–Condon approximation—like the original boson-sampling approach—which prevents us from accounting for Herzberg–Teller effects. Even so, the peak positions and overall band profiles are captured with high accuracy.

By eliminating the need to compute the excited-state Hessian—long recognized as the principal bottleneck in vibronic-spectra calculations[11]— our scheme both relaxes input-data requirements and, thanks to its linear runtime scaling, extends readily to molecular systems with hundreds of vibrational modes, replacing exponentially scaling eigenstate enumeration within the linear-coupling model. Photonic demonstrations that compensate optical losses achieve fidelities > 0.999 with superconducting-nanowire single-photon detectors and ≈ 0.99 with avalanche photodiodes, establishing a practical benchmark for detector technologies, while the same algorithm runs efficiently on conventional desktops and laptops, giving spectroscopists a broadly accessible, new tool for Franck–Condon profile prediction.

## Methods

### Reference spectrum

We employ the same fundamental physical approximations for both the time-independent sum-over-states (SOS) method and our sampling-based approach (e.g., 0 K, vacuum, harmonic oscillator, Franck–Condon, and linear coupling model). By ensuring the same starting assumptions, we isolate any discrepancies from the computational strategies rather than the differing physical models. The SOS method, while straightforward, requires calculating the energy and probability of every possible vibrational state. Because a harmonic oscillator has infinitely many levels, a cut-off must be introduced to cap the highest excitation level. Even then, the number of states grows exponentially, often making this approach prohibitively large, especially if many vibrational modes are retained. The spectra we call reference are generated with precisely these cut-offs. To enable a direct comparison, we apply the same cut-offs in our sampling-based method. Under these identical constraints—referred to as medium-sized systems—both approaches produce the similar results. In the optical experiment, the cut-off is enforced by using single-photon detectors that register any nonzero photon count as a single detection (i.e., maximum quantum number of one). Numerically, any random number exceeding the cut-off is capped at that value. After validating that both approaches match under these imposed limits (see Fig.4-5), we remove these artificial constraints in our sampling method and compare the outcome directly with experimental data (see Fig. 6). This final step is essential because as the system size grows, enumerating states becomes infeasible, making experimental comparison an indispensable alternative for verifying accuracy.

## Huang–Rhys factors

The calculation of Huang–Rhys factors generally requires the nuclear Hessians of both the $S_0$ and $S_1$ electronic states. However, for rigid molecules such as acenes, one of the most widely used computational approaches is the Linear Coupling Model (LCM)[22]. Within this approximation, the Duschinsky effect is neglected, meaning that the vibrational Hessians are assumed to be identical in both $S_0$ and $S_1$ states. The normal coordinates of the two states differ only by a displacement vector

$$\Delta Q_i = \frac{G_i}{\omega_i}$$

where $G_i$ is the gradient of $S_1$ along the ith normal mode at the optimized $S_0$ geometry, and $\omega_i$ is the vibrational frequency.

The Huang–Rhys factor for each mode is calculated as:

$$S_i = \frac{\omega_i \Delta Q_i^2}{2\hbar}$$

Geometry optimization for the ground-state singlet ($S_0$) was performed using density functional theory[37]. The energy gradients of the first excited singlet state ($S_1$) were calculated using the time-dependent DFT (TDDFT) level[38] at the optimized $S_0$ geometry. The calculations employed the B3LYP exchange–correlation functional[39,40] and the def2-TZVP basis set[41]. All computations were carried out using the TURBOMOLE software package[42–44].

# References

1. Gross, M. *et al.* Improving the performance of doped π-conjugated polymers for use in organic light-emitting diodes. *Nature* **405**, 661–665 (2000).

2. Hachmann, J. *et al.* The Harvard Clean Energy Project: Large-Scale Computational Screening and Design of Organic Photovoltaics on the World Community Grid. *J Phys Chem Lett* **2**, 2241–2251 (2011).

3. Lavis, L. D. & Raines, R. T. Bright Ideas for Chemical Biology. *ACS Chem Biol* **3**, 142–155 (2008).

4. Franck, J. & Dymond, E. G. Elementary processes of photochemical reactions. *Transactions of the Faraday Society* **21**, 536 (1926).

5. Condon, E. A Theory of Intensity Distribution in Band Systems. *Physical Review* **28**, 1182–1201 (1926).

6. Sharp, T. E. & Rosenstock, H. M. Franck—Condon Factors for Polyatomic Molecules. *J Chem Phys* **41**, 3453–3463 (1964).

7. Doktorov, E. V., Malkin, I. A. & Man'ko, V. I. Dynamical symmetry of vibronic transitions in polyatomic molecules and the Franck-Condon principle. *J Mol Spectrosc* **64**, 302–326 (1977).

8. Ruhoff, P. T. & Ratner, M. A. Algorithms for computing Franck-Condon overlap integrals. *Int J Quantum Chem* **77**, 383–392 (2000).

9.  Santoro, F., Lami, A., Improta, R. & Barone, V. Effective method to compute vibrationally resolved optical spectra of large molecules at finite temperature in the gas phase and in solution. *J Chem Phys* **126**, (2007).

10. Huh, J. & Berger, R. Application of time-independent cumulant expansion to calculation of Franck–Condon profiles for large molecular systems. *Faraday Discuss* **150**, 363 (2011).

11. Baiardi, A., Bloino, J. & Barone, V. General Time Dependent Approach to Vibronic Spectroscopy Including Franck–Condon, Herzberg–Teller, and Duschinsky Effects. *J Chem Theory Comput* **9**, 4097–4115 (2013).

12. Jankowiak, H.-C., Stuber, J. L. & Berger, R. Vibronic transitions in large molecular systems: Rigorous prescreening conditions for Franck-Condon factors. *J Chem Phys* **127**, (2007).

13. Petrenko, T. & Neese, F. Efficient and automatic calculation of optical band shapes and resonance Raman spectra for larger molecules within the independent mode displaced harmonic oscillator model. *J Chem Phys* **137**, (2012).

14. Huh, J., Guerreschi, G. G., Peropadre, B., McClean, J. R. & Aspuru-Guzik, A. Boson sampling for molecular vibronic spectra. *Nat Photonics* **9**, 615–620 (2015).

15. F. Duschinsky. On the Interpretation of Electronic Spectra of Polyatomic Molecules. I. The Franck–Condon Principle. *Acta Physicochim. URSS* **7** 551–556 (1937).

16. Wang, C. S. *et al.* Efficient Multiphoton Sampling of Molecular Vibronic Spectra on a Superconducting Bosonic Processor. *Phys Rev X* **10**, (2020).

17. Shen, Y. *et al.* Quantum optical emulation of molecular vibronic spectroscopy using a trapped-ion device. *Chem Sci* **9**, 836–840 (2018).

18. MacDonell, R. J. *et al.* Predicting molecular vibronic spectra using time-domain analog quantum simulation. *Chem Sci* **14**, 9439–9451 (2023).

19. Clements, W. R. *et al.* Approximating vibronic spectroscopy with imperfect quantum optics. *Journal of Physics B: Atomic, Molecular and Optical Physics* **51**, (2018).

20. Arrazola, J. M. *et al.* Quantum circuits with many photons on a programmable nanophotonic chip. *Nature* **591**, 54–60 (2021).

21. Zhu, H. H. *et al.* Large-scale photonic network with squeezed vacuum states for molecular vibronic spectroscopy. *Nat Commun* **15**, (2024).

22. Macak, P., Luo, Y. & Ågren, H. Simulations of vibronic profiles in two-photon absorption. *Chem Phys Lett* **330**, 447–456 (2000).

23. Gierschner, J., Mack, H.-G., Lüer, L. & Oelkrug, D. Fluorescence and absorption spectra of oligophenylenevinylenes: Vibronic coupling, band shapes, and solvatochromism. *J Chem Phys* **116**, 8596–8609 (2002).

24. Gierschner, J. *et al.* Optical spectra of oligothiophenes: vibronic states, torsional motions, and solvent shifts. *Synth Met* **138**, 311–315 (2003).

25. Dierksen, M. & Grimme, S. Density functional calculations of the vibronic structure of electronic absorption spectra. *J Chem Phys* **120**, 3544–3554 (2004).

26. Prampolini, G. *et al.* Computational Design, Synthesis, and Mechanochromic Properties of New Thiophene-Based π-Conjugated Chromophores. *Chemistry – A European Journal* **19**, 1996–2004 (2013).

27. Atkins, P. W. & Friedman, R. S. *Molecular Quantum Mechanics*. (Oxford University Press, Oxford, 2011).

28. Huang, K. & Rhys, A. Theory of light absorption and non-radiative transitions in *F*-centres. *Proc R Soc Lond A Math Phys Sci* **204**, 406–423 (1950).

29. Kahl, O. *et al.* Waveguide integrated superconducting single-photon detectors with high internal quantum efficiency at telecom wavelengths. *Sci Rep* **5**, 10941 (2015).

30. Kovalyuk, V. *et al.* Absorption engineering of NbN nanowires deposited on silicon nitride nanophotonic circuits. *Opt Express* **21**, 22683 (2013).

31. Valiev, R. R., Cherepanov, V. N., Baryshnikov, G. V. & Sundholm, D. First-principles method for calculating the rate constants of internal-conversion and intersystem-crossing transitions. *Physical Chemistry Chemical Physics* **20**, 6121–6133 (2018).

32. Ferguson, J., Reeves, L. W. & Schneider, W. G. Vapor absorption spectra and oscillator strengths of naphthalene, anthracene, and pyrene. *Can J Chem* **35**, 1117–1136 (1957).

33. Maeda, H., Maeda, T. & Mizuno, K. Absorption and Fluorescence Spectroscopic Properties of 1- and 1,4-Silyl-Substituted Naphthalene Derivatives. *Molecules* **17**, 5108–5125 (2012).

34. Benkyi, I., Tapavicza, E., Fliegl, H. & Sundholm, D. Calculation of vibrationally resolved absorption spectra of acenes and pyrene. *Physical Chemistry Chemical Physics* **21**, 21094–21103 (2019).

35. Herzberg, G. *Molecular Spectra and Molecular Structure. Vol. III: Electronic Spectra and Electronic Structure of Polyatomic Molecules.* (Van Nostrand Reinhold, New York, 1966).

36. Avila Ferrer, F. J. & Santoro, F. Comparison of vertical and adiabatic harmonic approaches for the calculation of the vibrational structure of electronic spectra. *Physical Chemistry Chemical Physics* **14**, 13549 (2012).

37. Parr, R. G. & Yang, W. *Density-Functional Theory of Atoms and Molecules*. (Oxford University Press, New York, 1989).

38. Runge, E. & Gross, E. K. U. Density-Functional Theory for Time-Dependent Systems. *Phys Rev Lett* **52**, 997–1000 (1984).

39. Becke, A. D. Density-functional thermochemistry. III. The role of exact exchange. *J Chem Phys* **98**, 5648–5652 (1993).

40. Lee, C., Yang, W. & Parr, R. G. Development of the Colle-Salvetti correlation-energy formula into a functional of the electron density. *Phys Rev B* **37**, 785–789 (1988).

41. Weigend, F. & Ahlrichs, R. Balanced basis sets of split valence, triple zeta valence and quadruple zeta valence quality for H to Rn: Design and assessment of accuracy. *Physical Chemistry Chemical Physics* **7**, 3297 (2005).

42. Ahlrichs, R., Bär, M., Häser, M., Horn, H. & Kölmel, C. Electronic structure calculations on workstation computers: The program system turbomole. *Chem Phys Lett* **162**, 165–169 (1989).

43. Franzke, Y. J. *et al.* TURBOMOLE: Today and Tomorrow. *J Chem Theory Comput* **19**, 6859–6890 (2023).

44. Balasubramani, S. G. *et al.* TURBOMOLE: Modular program suite for *ab initio* quantum-chemical and condensed-matter simulations. *J Chem Phys* **152**, (2020).

## Acknowledgements

This work was supported by the PhoQuant project and by the Academy of Finland (grant no. 340583). The authors also acknowledge CSC – IT Center for Science, Finland, for providing the computational resources used in this study.

## Author contributions

- Conceptualization: IK

- Methodology: IK

- Design and fabrication of SNSPD chip: IK, OP

- Computation of Huang–Rhys factors and vibrational frequencies: RTN, CÜ, VVR

- Experiments using SNSPD and SPAD detectors: IK, SF, RP

- PC-based simulations and fidelity analysis: IK

- Project administration: WP, SF

- Funding acquisition: WP

- Supervision: WP, SF

- Writing – original draft: IK

- Writing – review & editing: All authors

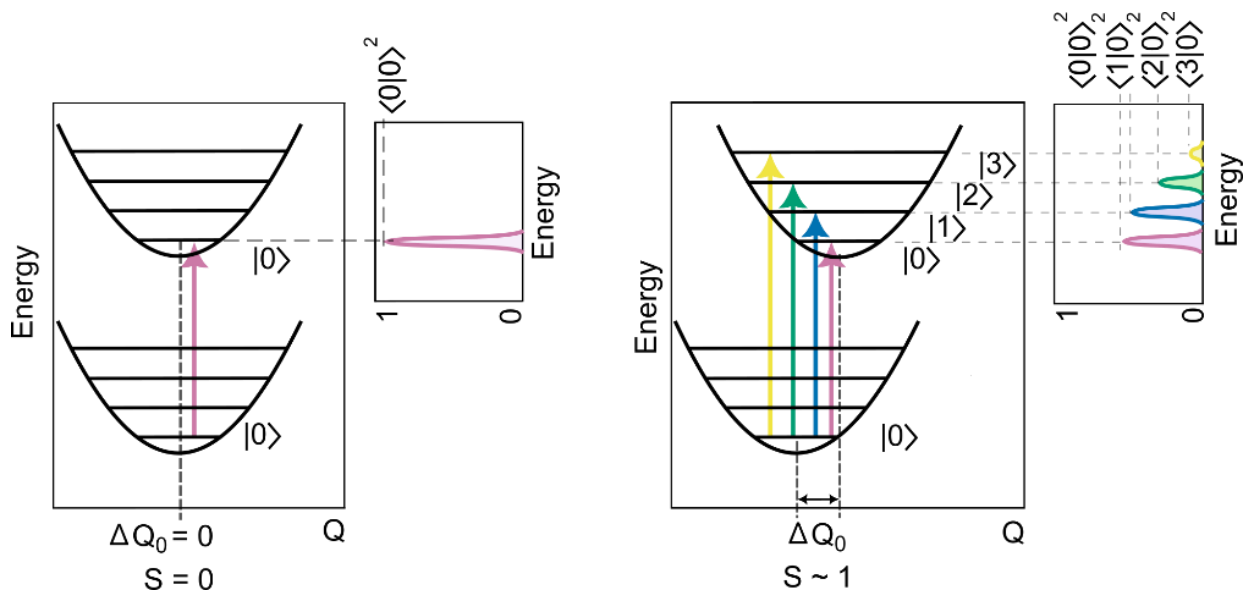

**Figure 1: Jablonski diagram (left) and Franck–Condon profile (right) for a one mode system in the linear coupling model (LCM).** In the LCM, the initial and final electronic states of the vibronic transition are each described by harmonic oscillators with the same frequency. A key parameter that determines the shape of the Franck–Condon profile is the shift $\Delta Q_0$ in the equilibrium position of these oscillators. This shift is directly related to the average number of phonons $S$ that are created during the transition, since $S$ is proportional to $\Delta Q_0^2$. In the figure on the left, the electronic transition does not alter the equilibrium position, so $\Delta Q_0 = 0$. Consequently, the average number of phonons created is zero, leading to the absence of vibronic sidebands; we see only a zero-phonon line in the Franck–Condon profile. In contrast, the figure on the right shows a case where the electronic transition induces a noticeable shift in the equilibrium position—about enough to create one phonon on average. As a result, a rich vibronic structure appears in the Franck–Condon profile.

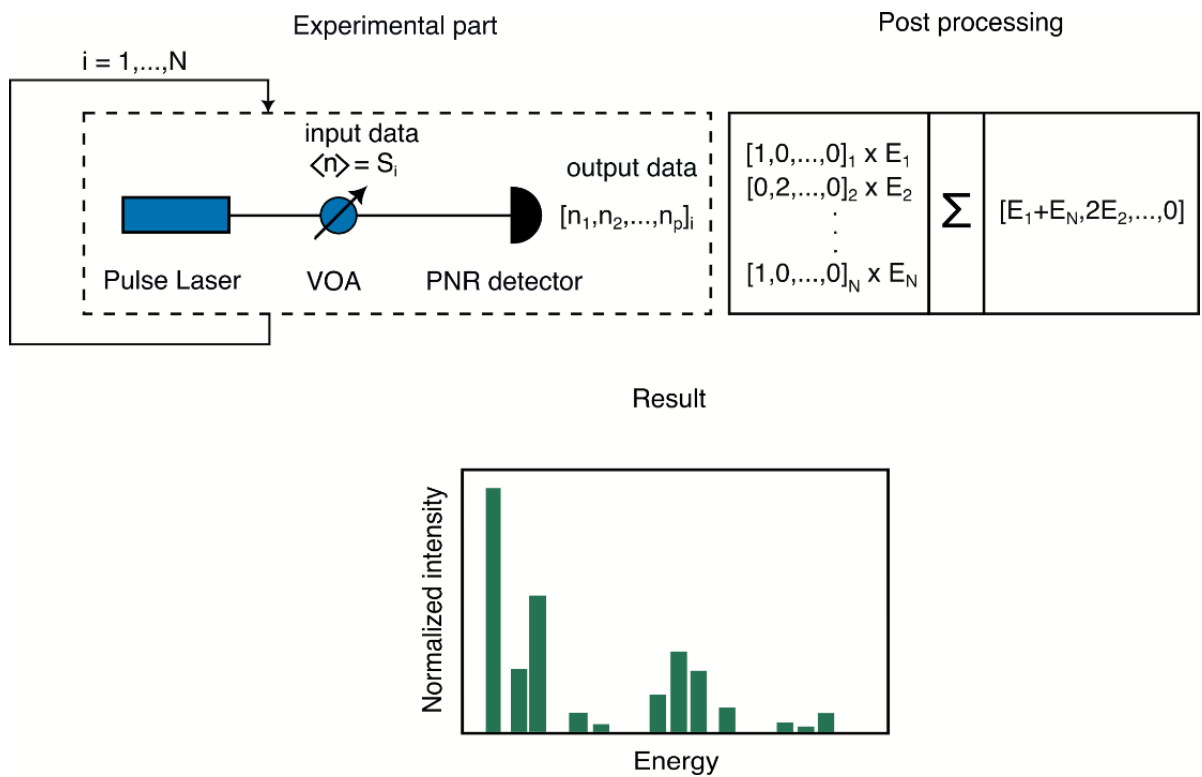

**Figure 2: Algorithm for calculating the Franck-Condon profile within the linear coupling model (LCM) with an optical implementation.** The algorithm consists of two main steps. In the first step, we generate Poisson-distributed integers for a chosen average value. In the optical setup, this is done by encoding the average phonon number of each vibrational mode into the average photon number of a pulsed laser, adjusting it with a variable optical attenuator (VOA). A photon-number-resolving (PNR) detector records the outcome of each pulse into an array, producing one array per vibrational mode. Then we repeat this procedure for all modes of the molecule. In the second step (post-processing), each array is multiplied by the vibrational energy of its corresponding mode. Summing these arrays element-wise gives a final array, whose normalized histogram reveals the molecule's Franck–Condon profile.

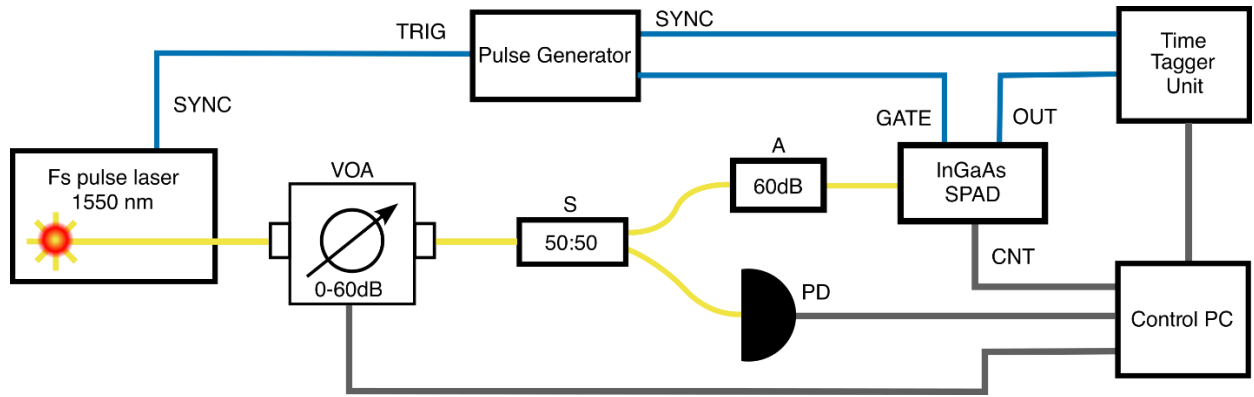

**Figure 3: Schematic of the optical Setup.** A pulsed laser is attenuated by a variable optical attenuator (VOA) and split by a 50:50 coupler. One output is monitored by a photodiode to track the laser power; the other output passes through a fixed attenuation before detection by a gated single-photon avalanche diode (SPAD). A pulse generator, triggered by the laser, provides synchronized gating signals to the SPAD. The laser trigger and SPAD events are recorded by a time tagger, enabling photon-number analysis from the fraction of null detections. Alternatively, the same optical layout is used with a superconducting nanowire single-photon detector (SNSPD), requiring no gating and offering negligible afterpulsing, with identical data processing in the time tagger.

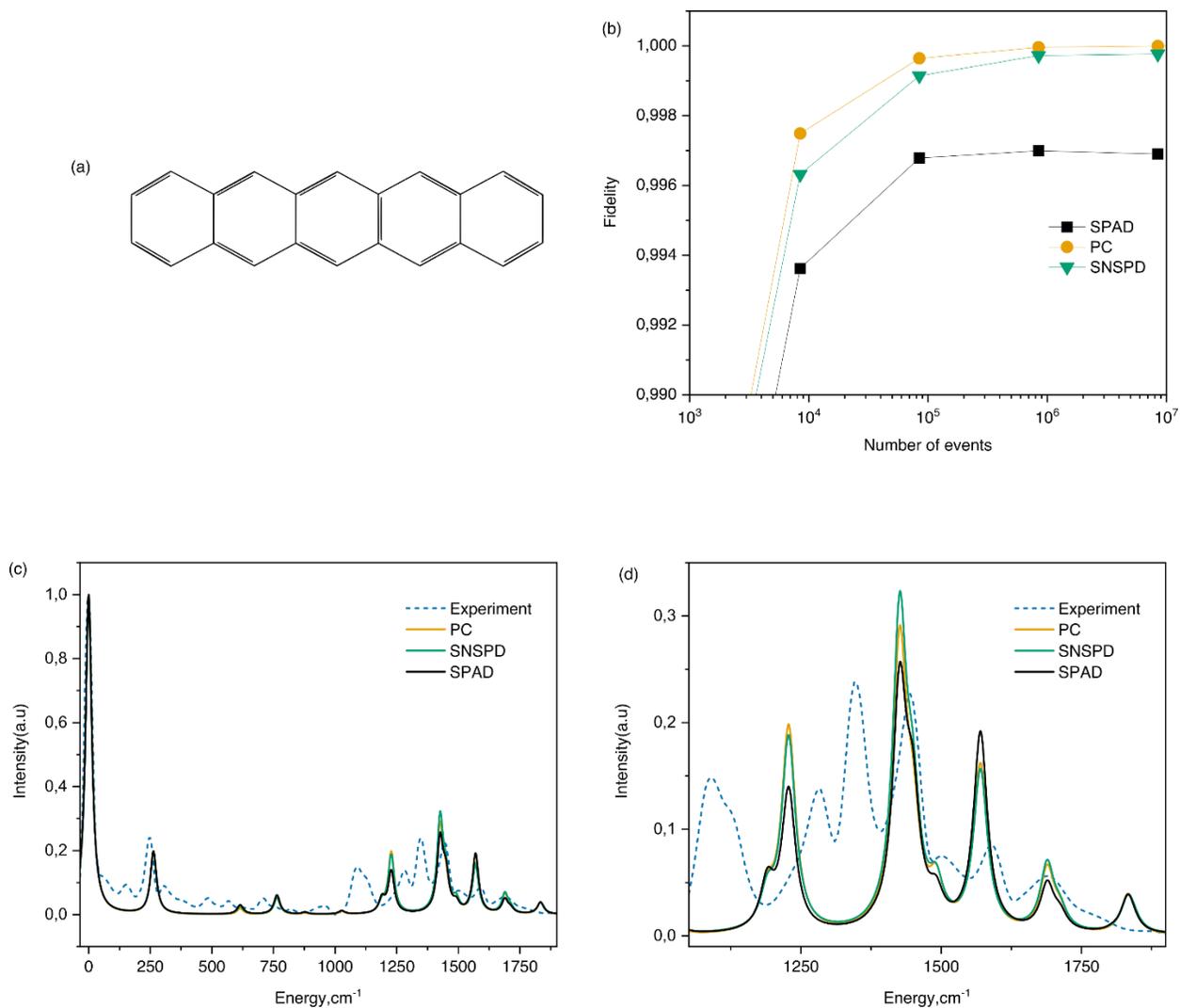

**Figure 4: Results for the pentacene molecule on different hardware platforms.** a) The molecular structure of Pentacene molecule. Since the pentacene molecule consists of 36 atoms, it has 102 vibrational modes. Most of them do not contribute to the vibronic progression, so we consider the 8 modes with the largest values of the Huang-Rhys Factors. (b) Fidelity values vs number of counts for different hardware implementation of proposed algorithm. In all three cases, with the accumulation of statistics of the number of events of more than 10 thousand, we have a fidelity value of more than 99%. c) absorption spectrum $S_0 \rightarrow S_1$ transition of Pentacene: experimental data and simulated data and reference spectrum

calculated analytically. All spectrums are normalized to the intensity of the 0-0 line. d) Enlarged fragment of the graph c.

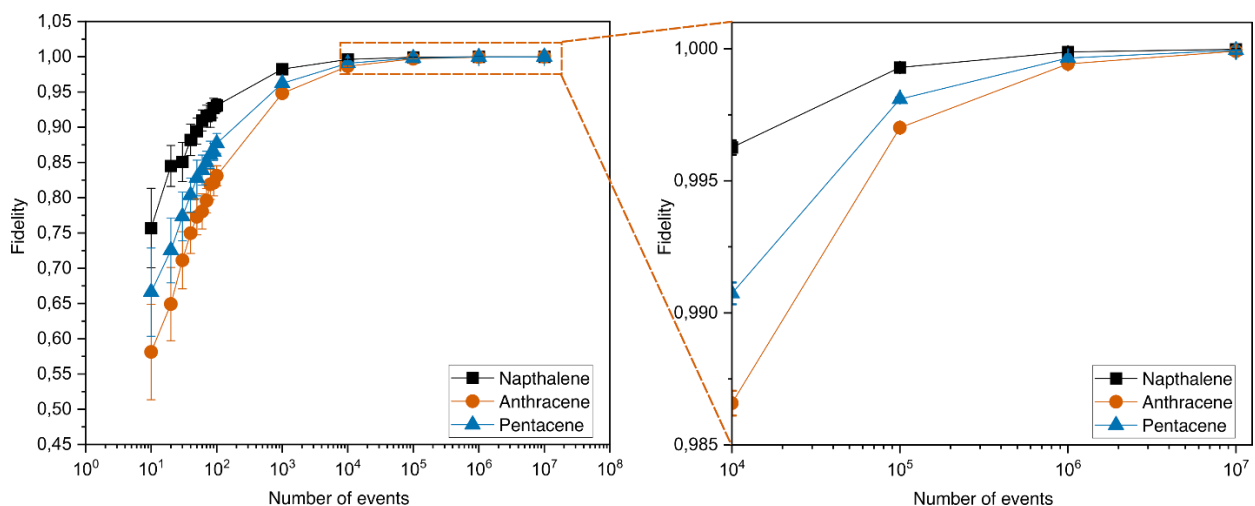

**Figure 5. Convergence of the sampling-based algorithm for molecules with higher vibrational excitation and larger mode counts.** Left: Fidelity F between the spectrum obtained with the present algorithm and the exact sum-over-states reference is shown as a function of the number of detected events for naphthalene (K = 3, 9 modes), anthracene (K = 3, 12 modes) and pentacene (K = 1, 18 modes). Right: Magnified view of the high-fidelity region (dashed frame) illustrates that all three systems reach F > 0.99 once the sample size exceeds $10^5$ events. The similar convergence trends confirm that the algorithm retains accuracy when the maximum vibrational quantum number K increases beyond unity and when the number of included normal modes is more than doubled. Error bars denote one standard deviation over 30 independent runs.

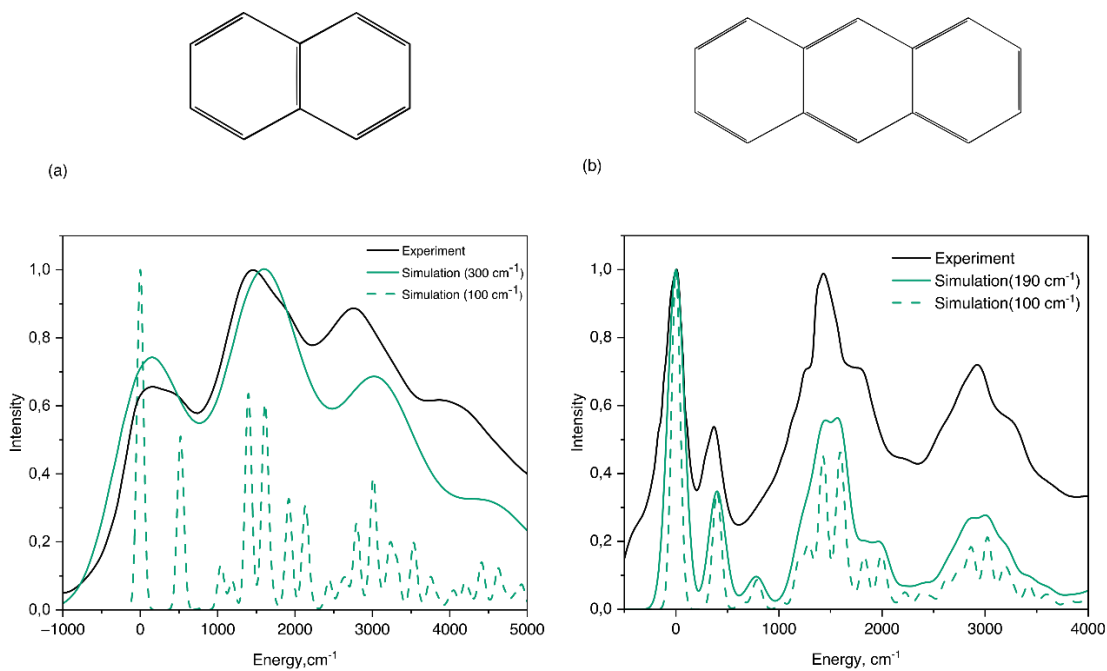

**Figure 6: Comparison of spectra obtained using the proposed method with absorption spectra measured at room temperature for naphthalene and anthracene.** a) top: The molecular structure of naphthalene molecule, bottom: absorption spectrum $S_0 \rightarrow S_1$ transition of Naphthalene: experimental data (black line) and simulated data (green lines). All spectra are normalized to the maximum value of intensity b) top: The molecular structure of the anthracene molecule, bottom: absorption spectrum of the $S_0 \rightarrow S_1$ transition of anthracene: experimental data (black line) and simulated data (green lines). All spectra are normalized to the maximum intensity value and expanded using Gaussian functions with a full width at half maximum: $100 cm^{-1}$ (dashed line), $300 cm^{-1}$, $190 cm^{-1}$ (solid green line on a) and b) respectively)